\documentstyle[11pt,cite,epsf]{article}

\renewcommand{\thefootnote}{\fnsymbol{footnote}}    
    
\begin{document}    
    

\begin{titlepage}      

\begin{center}           
\hfill    KIAS P05028\phantom{i}
\\ 
\hfill SNUTP 05004
\\          
\hfill  UOSTP 05013
\\
\hfill  {\tt hep-th/0504098}        
\vspace{2cm}

{\Large \bf 1/2 BPS Geometries of M2 
Giant Gravitons
}

\vspace{1.0cm}           
{\large           
{\bf Dongsu Bak},$\!^a$ {\bf Sanjay Siwach}$\,^{a,b}$  and 
{\bf Ho-Ung Yee}$\,^c$            
}           
           
\vspace{0.6cm}

{\it  $^a$          
Physics Department, University of Seoul, Seoul 130-743, Korea           
}           
\vskip 0.3cm        
{\it  $^b$          
School of Physics, Seoul National university, Seoul 151-743, Korea           
}           
\vskip 0.4cm          
\centerline{\it $\,^{c}$ School of Physics, Korea Institute          
for Advanced Study}          
\centerline{
\it Cheongriangri-Dong, Dongdaemun-Gu, Seoul 130-012, Korea}

({\tt dsbak@mach.uos.ac.kr, sksiwach@phya.snu.ac.kr, ho-ung.yee@kias.re.kr})

\end{center}           
\vspace{1.5cm}

We construct the general 1/2 BPS M2 giant graviton
solutions asymptotic to the eleven-dimensional maximally
supersymmetric plane wave background, based on the recent work of
Lin, Lunin and Maldacena. The solutions have  null singularity and
we argue that it is unavoidable
to have null singularity in the proposed framework,
although the solutions are still physically relevant. They involve
an arbitrary function $F(x)$ which is shown to have a
correspondence to the 1/2 BPS states of the BMN matrix model. A
detailed map between the  1/2 BPS states of both sides is worked
out. 

\vspace{2.5cm}           
\begin{center}           
\today           
\end{center}           
\end{titlepage}   




\newcommand{\sect}[1]{\section{#1}}
\newcommand{\subsect}[1]{\subsection{#1}}
\newcommand{\subsubsect}[1]{\subsubsection{#1}}

\newcommand{\be}{\begin{equation}}
\newcommand{\ee}{\end{equation}}
\newcommand{\bea}{\begin{eqnarray}}
\newcommand{\eea}{\end{eqnarray}}

\newcommand{\up}{\uparrow}
\newcommand{\down}{\downarrow}
\newcommand{\lr}{\leftrightarrow}

\newcommand{\la}{\langle}
\newcommand{\ra}{\rangle}
\newcommand{\da}{\dagger}
\newcommand{\p}{\partial}
\newcommand{\ep}{\epsilon}
\newcommand{\varep}{\varepsilon}

\newcommand{\hi}{\hat{\i}}
\newcommand{\hj}{\hat{\j}}
\newcommand{\nn}{\nonumber}
\renewcommand{\thefootnote}{\fnsymbol{footnote}}

\def\del{\nabla}
\def\half{\frac{1}{2}}


\def\CM{{\mathcal{M}}}
\def\CH{{\mathcal{H}}}
\def\CN{{\mathcal{N}}}
\def\CF{{\mathcal{F}}}
\def\CP{{\mathcal{P}}}
\def\CQ{{\mathcal{Q}}}
\def\CD{{\mathcal{D}}}
\def\CO{{\mathcal{O}}}

\def\IB{\relax\hbox{$\inbar\kern-.3em{\rm B}$}}
\def\IC{\relax\hbox{$\inbar\kern-.3em{\rm C}$}}
\def\ID{\relax\hbox{$\inbar\kern-.3em{\rm D}$}}
\def\IE{\relax\hbox{$\inbar\kern-.3em{\rm E}$}}
\def\IF{\relax\hbox{$\inbar\kern-.3em{\rm F}$}}
\def\IG{\relax\hbox{$\inbar\kern-.3em{\rm G}$}}
\def\IGa{\relax\hbox{${\rm I}\kern-.18em\Gamma$}}
\def\IH{\relax{\rm I\kern-.18em H}}
\def\IK{\relax{\rm I\kern-.18em K}}
\def\IL{\relax{\rm I\kern-.18em L}}
\def\IP{\relax{\rm I\kern-.18em P}}
\def\IR{\relax{\rm I\kern-.18em R}}
\def\IZ{\relax{\rm Z\kern-.5em Z}}


\def\apr{\alpha'}
\def\str{{str}}
\def\tr{\hbox{tr}}
\def\lstr{\ell_\str}
\def\gstr{g_\str}
\def\Mstr{M_\str}
\def\lpl{\ell_{pl}}
\def\Mpl{M_{pl}}




\setcounter{footnote}{0}


\section{Introduction}

Recently, comprehensive works on $1/2$ BPS geometries in the type
IIB and 11D supergravities \cite{lin}, as well as in the six
dimensional supergravity have been done
\cite{Lunin:2001jy,Martelli:2004xq,Liu:2004hy}. (For other related
works please see \cite{Bena:2004qv}.) They are supposed to
describe geometries corresponding to $1/ 2$ BPS states of  dual
CFT in the framework of AdS/CFT correspondence \cite{malda, bmn}.
The story in the case of type IIB turns out to be much
interesting, because in this case we know much information on both
the geometry side \cite{lin,mst} and the corresponding CFT side
\cite{ber}. The two dimensional phase space of the emergent
fermion picture in the $1/ 2$ BPS sector of $\CN=4$ SYM theory is
geometrically realized as a surface of specific boundary
conditions. Detailed matching between $1/2$ BPS operators in
$\CN=4$ SYM and the geometries of giant gravitons, in a way
suggested by \cite{ber,Caldarelli:2004ig}, provides another
convincing example of AdS/CFT correspondence. It was pointed out
that it is even possible to identify the fermion system directly
in the supergravity side \cite{Mandal:2005wv}.

Contrary to $1/ 2$ BPS geometries of $AdS_5\times S^5$ in type
IIB, where we only need to solve a linear Laplace equation with a
specified boundary condition, the $1/2$ BPS geometries in the
M-theory are much harder to find because we have to solve a
continuum version of nonlinear Toda-like equation \cite{lin}. The
fact that we again have a boundary plane divided into  regions of
two different kinds of boundary conditions seems to be similar to
the type IIB case, but the nature and even the existence of smooth
solutions are not very clear yet. However, for the geometries in
the background 11-dimensional plane-wave, in which case there is a
translational symmetry along one direction on the boundary plane,
it is possible to map the non-linear equation to the linear
Laplace equation \cite{lin,Ward}. In this work, we analyze this linear
equation carefully to find solutions of physical relevance.

The solutions we provide in this work depend on an arbitrary
function $F(x)$, which we interpret as the number density of
$1/2$ BPS spherical M2-branes of radius $x$ in the transverse
$R^3$ space of the 11D plane wave background. This identification
will be obtained from comparing the charges of the geometries with
those of $1/2$ BPS solutions in the BMN matrix model.
However, we find that the geometries of our interest have null
conical singularity. In fact, we will try to give a convincing
argument that this type of singularity is  inevitable in the
proposed ansatz of Lin, Lunin and Maldacena (LLM). We leave it as
an open question to discuss the acceptance of this type of
singularity. Except this singularity, the resulting geometry seems
to have a nice correspondence to the BMN matrix model expectations
\cite{bmn} .

After a brief review on the $1/ 2$ BPS geometries in M-theory in the next section,
we construct the explicit solutions of $1/ 2$ BPS geometries in the
plane-wave background in section 3, followed by explicit examples and
analysis in section 4. In section 5, we discuss the singularity
and argue that it seems impossible to avoid the singularity in the
given ansatz. We compute the charges of the geometries and finally
compare them with the BMN matrix model in section 6 and 7.

\section{A Review of LLM Construction in M-theory}

In this section we review the construction of 
1/2 BPS geometries in M-theory following Lin, Lunin and Maldacena
(LLM)~\cite{lin}.
M-theory or the eleven dimensional supergravity in the low energy
limit contains two familiar 1/2 BPS objects, M2 and M5 branes. M2
(M5) brane is electrically (magnetically) charged with respect to
the three from gauge potential of the eleven dimensional
supergravity. The corresponding supergravity solutions interpolate
between flat $\CM_{11}$ and $AdS_4\times S^7$ ($AdS_7\times S^4$)
geometries for M2 (M5) branes, which are known to be maximally
supersymmetric solutions of M-theory. In addition to this, one
additional background is known which is also maximally
supersymmetric. This corresponds to the plane wave solution
 obtained from  $AdS_4\times S^7$
or $AdS_7\times S^4$ by taking the Penrose limit \cite{Blau}.

Motivated by AdS/CFT correspondence \cite{malda} and its
extension to plane-wave/CFT correspondence \cite{bmn}, one is
interested in the BPS states of the dual CFT obeying $\Delta = J$,
where $\Delta$ is the conformal dimension  of the CFT fields and
$J$ corresponds to the $U(1)$ charge in the $R$ symmetry group.
Let us consider 1/2 BPS geometries in AdS$_7\times S^4$. These are
associated
to the chiral primaries of the (2,0) 
theory. The chiral primaries of (2,0) theory can be described in
terms of two dimensional Young diagrams. These states preserve
half of the supersymmetry and are invariant under the symmetry group,
 $SO(3)\times SO(6)\times R$ of the (2,0) theory on $R_t\times S^5$. In the dual
supergravity description, one is interested in the half BPS
solutions with this symmetry.

The most general ansatz for the eleven dimensional background consistent with
$SO(3)\times SO(6)$ symmetry can be
written as \cite{lin}
\bea
\label{ansatz}
ds^2 &  = & e^{2\lambda}\left(\frac{1}{m^2}d\Omega^2_5 + e^{2A}
d\Omega^2_2 + ds^2_4\right) \cr
&\cr
G_{(4)} & = & G_{\mu_1\mu_2\mu_3\mu_4}\, dx^{\mu_1}
\wedge dx^{\mu_2} \wedge dx^{\mu_3}\wedge dx^{\mu_4} +
F_{\mu_1 \mu_2}\,\, dx^{\mu_1} \wedge dx^{\mu_2} \wedge d^2\Omega^2_2
\eea
where $\Omega^2_2$  and $\Omega^2_5$ being the line elements
of the unit two sphere and five sphere respectively and $G_{(4)}$ being the field
strength of the three form gauge potential of eleven
dimensional supergravity. The Greek indices run over
 $1,..,4$ and refer to the four dimensional metric $ds^2_4$.
$\lambda$ and $A$ are functions of all
the coordinates and $m^2$ is some fixed parameter independent of the
coordinates.

The number of supersymmetries preserved by a background is in one
to one correspondence with the number of covariantly constant
Killing spinors admitted by the background. To find solutions
which preserve some fraction of the supersymmetry one has to solve
the Killing spinor equation \cite{Gauntlett:2002fz},
\be
\nabla_m \xi +{1\over 288} \left(
\Gamma_m^{npqr}- 8\delta_m^n \Gamma^{pqr} \right) G_{(4)npqr}\xi =
0.
\ee
One can solve the above equation by decomposing the spinor, $\xi$
into spinors of $ds^2_4$ and seven dimensional spinors and perform
the dimensional reduction on $S^5$. Further reduction on $S^2$
gives the equations for the 4-dimensional spinor. One can fix the
form of the metric by working out the properties of the Killing spinor
bilinears and using Fierz identities.
Defining a new coordinate, $y$ in terms of the bilinears, the metric
components can be written in terms of a single function, $D$ of
spatial coordinates of the four dimensional metric $ds^2_4$ (for the detailed
discussion of the reduction procedure and fixing the form of
metric, we refer to the appendix of Ref. \cite{lin}) \bea
\label{met} ds^2 & = & -4 e^{2\lambda}\left(1 + y^2
e^{-6\lambda}\right) \left(dt + V_idx^i\right)^2 +
\frac{e^{-4\lambda}}{1 + y^2 e^{-6\lambda}}\left[dy^2 +
e^D\left(dx^2_1 + dx^2_2\right)\right]  \nn \\
&&+ 4 e^{2\lambda}d\Omega^2_5 + y^2 e^{-4\lambda}d\Omega^2_2 \cr
&\cr e^{-6\lambda} & = & \frac{\p_y D}{y(1 - y\p_y D)} \cr &\cr
V_i & = & \half \ep_{ij}\p_j D\quad, \eea where $i,j=1,2$. We have
not written the expression for the four form $G_{(4)}$, as we
shall not need it. It is also determined completely from the
knowledge of the function $D(y, x_1, x_2)$. The parameter $m$ in
the equation (\ref{ansatz}) has been set to $m=\half$. The
function $D(y, x_1, x_2)$ satisfies the three dimensional Toda
equation \be \label{toda} \left(\p^2_1 + \p^2_2\right) D + \p^2_y
e^D = 0. \ee

It can be seen from the metric (\ref{met}) that the coordinate
$y$ is related to the radii of the two sphere, $R_2$ and of the
five sphere, $R_5$ by the relation $y = R_2 R^2_5/4$. At $y = 0$
either two-sphere or five-sphere shrinks to zero size. The
boundary conditions at $y = 0$ are required to be such that the
geometry remains non singular when either of the spheres shrinks.
In the case where two-sphere shrinks while the radius of
five-sphere remains constant, one finds $\p_y D = 0$ at $y = 0$
and $D$ is independent of $y$. In the other case where
five-sphere shrinks to zero size while the radius of two-sphere remains
constant, we have $D\sim \log y$ at $y = 0$. Thus, one can write
the boundary conditions at $y=0$ as \bea \label{bc} \p_y D & = & 0,
\>\>\> D =  {\rm finite} , \ \ \ {\rm when}\  S^2 \  {\rm shrinks}
\cr &\cr e^D  & \sim &  y, \ \ \ \ {\rm when} \ S^5 \ {\rm
shrinks} \eea We shall refer the above boundary conditions as
`finite type' and `linear type' respectively as the function $e^D$
becomes finite or goes linearly with $y$. We shall be interested
in solutions of the Toda equation (\ref{toda}) satisfying the above
boundary conditions. In general the explicit solutions to Toda
equations are not known, but in the particular case when there is
an extra spatial isometry (say $x_1$ in our case) the Toda
equation (\ref{toda}) reduces to \be \label{toda1} \p^2_2 D +
\p^2_y e^D = 0. \ee By a change of variable\footnote{Our notation
differs from Ref.~\cite{lin} by $x_2\rightarrow -x_2$.} \be
\label{defyax} e^D = \rho^2, \>\>\>\>\>\>\>\>\> y = \rho\p_\rho V,
\>\>\>\>\>\>\>\>\> x_2 = -\p_\eta V \ee the Toda equation
(\ref{toda1}) can be mapped to the Laplace equation in three
dimensions\cite{Ward}
\be
\label{lap} \frac{1}{\rho}\p_\rho\left(\rho\p_\rho V\right) +
\p^2_\eta V = 0. \ee Therefore, the task of solving the Toda
equation (\ref{toda1}) is reduced to solving
 the Laplace equation in the new coordinates subject to the
boundary conditions (\ref{bc}).

\section{M-Theory Giant Gravitons}

In this section we are interested in the solutions of the Laplace
equation (\ref{lap}) which correspond to giant gravitons in
M-theory. (See Ref.\cite{Lee:2004kv} for a linearized solution for
M-theory giant graviton.) In particular we shall be interested in
the solutions which are obtained from superposition of the
solutions of the Laplace equation.
The plane wave background is
given by a solution of the Laplace equation \cite{lin}; \be V_0 =
\rho^2 \eta - \frac{2}{3}\eta^3\quad.\ee As a warm-up exercise,
let us discuss the boundary condition for this solution
explicitly. The metric for this solution can be written
as \be \label{ppmet} ds^2 =  -4 (\rho^2 + 4 \eta^2) dt^2 - 4
dx_1dt  +
4 \left(d\rho^2 + d\eta^2\right)   \nn \\
+ 4 \rho^2 d\Omega^2_5 + 4 \eta^2 d\Omega^2_2\quad. \ee We see
that $\rho$ and $\eta$ are  nothing but the radii of the
five-sphere and two-sphere respectively. We are interested in the
issue of boundary conditions at $y = 0$ in terms of the new
coordinates $(\rho, \eta)$. They are related to the previous $(y,
x_2)$ coordinates by \be \label{yaxppwave} y   =  2 \rho^2 \eta
\,, \hspace{1.5cm} x_2 = -(\rho^2 -2\eta^2) \quad. \ee The
boundary at $y = 0$ in the $(\rho, \eta)$-plane corresponds to either
$\rho \rightarrow 0$ ($S^5$ shrinking), while keeping $\eta$
(radius of two sphere) finite or $\eta \rightarrow 0$ ($S^2$
shrinking), while keeping $\rho$ (radius of five sphere) finite.
When the five sphere shrinks one can see that the boundary conditions at $y
= 0$ are of the `linear type' that is $e^D \sim y$. In the other case
when the two sphere shrinks, the boundary conditions at $y = 0$ are of
the `finite type' that is $e^D \sim {\rm finite}$.

Now we shall construct solutions which are superpositions of the
solutions of Laplace equation (\ref{lap}) upon the plane wave
background. Later we shall identify these solutions as half BPS
states of BMN matrix model \cite{bmn}.  Solutions to the Laplace
equation have several types. There are
 polynomial
type solutions like the case of the above plane wave solution. To
recover the plane-wave background asymptotically, they should be
absent. Other class of solutions are the superposition of \be K_0
(k\rho)\, e^{ik\eta}\,, \ \ \ \ I_0 (k\rho)\, e^{ik\eta}\,, \ee or
\be J_0 (k\rho) e^{k\eta}\,, \ \ \ \,\, N_0 (k\rho) e^{
k\eta}\,, \ee in terms of the Bessel functions. After careful
analysis of these types of solutions, it turns out that the
relevant one for our application is the solutions involving $K_0(k\rho)$,
the zeroth order Bessel function of the imaginary argument, which
shows the logarithmic singularity at $\rho=0$.
For this choice, generic solutions to the Laplace equation
(\ref{lap}) can be written as \be V_1 =  2\int^{\infty}_{-\infty}
dk A(k) K_0(k\rho) e^{i k \eta}\,. \ee
Using the integral representation for the Bessel function $K_0(x)$
\be K_0(x)  =  \int^{\infty}_{0} dt \frac{\cos tx}{\sqrt{1 + t^2}}
 = {1\over 2} \int^{\infty}_{-\infty} dt \frac{e^{itx}}{\sqrt{1 + t^2}}
\quad,\ee the final form of the solution obtained from superposing
on the plane-wave background is \be V = V_0+V_1=\rho^2 \eta -
\frac{2}{3}\eta^3 + \int^{\infty}_{-\infty} dt \frac{F(\eta +
t\rho)}{\sqrt{1 + t^2}} \ee where the function $F(x)$ is the
inverse Fourier transform of $A(k)$. In fact, $V_1$ is the
electromagnetic potential of a line charge density $F(\eta)$
distributed along the $\eta$-axis. 

The other choices of polynomials does not lead to the asymptotically 
plane wave geometries. One may also show that the choices of $I_0$, 
$J_0$ and $N_0$ are not consistent with the asymptotically plane wave 
geometries with the desired boundary conditions. 
In section 5, we will actually
consider the most general acceptable form of $V_1$ for the
solutions of the Laplace equation, and compare with the properties 
of the solutions in this
section.


Using (\ref{defyax}), the expression for $y$ and $x_2$ may be
written as \bea \label{yax} y \,\, & = & 2 \rho^2 \eta  + \rho
\int^{\infty}_{-\infty} dt \frac{t}{\sqrt{1 + t^2}}\,  F'(\eta +
t\rho)  \cr &\cr x_2 & = & -\rho^2 +2\eta^2  -
\int^{\infty}_{-\infty} dt \frac{1}{\sqrt{1 + t^2}}\, F'(\eta +
t\rho) \label{master}\eea where prime denotes taking derivative
with respect to the argument.

In terms of the  coordinates ($\rho$, $\eta$), the expression for
metric can be written explicitly as: \bea \label{met1} ds^2 & = &
4 e^{-4\lambda} \left [-
W 
\left(dt + V_1dx_1\right)^2 +
\frac{\rho^2 }{4 W
} e^{6\lambda}(dx_1)^2 \right] +
\frac{K}{W
}e^{2\lambda} \left(d\rho^2 + d\eta^2\right)   \nn \\
&&+ 4 e^{2\lambda}d\Omega^2_5 + y^2 e^{-4\lambda} d\Omega^2_2  \cr
&\cr
e^{6\lambda}  & = & W
- y^2, \hspace{1.5cm} V_1  = \frac{1}{K}\frac{\p y}{\p
\eta}.\label{newmetric} \eea
where $W
= \frac{y \rho K}{2 \p_\rho y} $ and $K = \rho J$, $J$ being the
Jacobian of transformation from $(y, x_2)$ coordinates to $(\rho,
\eta)$ coordinates (see below). Unlike the case in terms of the
coordinates $(y,x_2)$, the solution here is no longer implicit.
Namely, all the metric components as well as the field strengths
can be written explicitly in terms of $F(x)$, though the actual
expression may not be much informative .

Let us discuss the boundary conditions for these solutions
carefully.
To get the right kind of boundary
conditions (\ref{bc}) for  giant gravitons, we  
restrict the function $F(x)$ to be an odd function. This choice
will be further explained in the later section.  In addition,
$F(x)$ should be nonnegative for $x\ge 0$.
(For the negative $F(x)$, the $y=0$ line stops somewhere in the $\eta$ axis.
In this case,  at the boundary
of $\eta$ axis excluded by the $y=0$ line, the radii of two and five
spheres become finite while $\rho$ goes to zero. For a smooth
geometry, one has to utilize $x_1$ circle but there is no way
to match the period of the $x_1$ circle except for some special
cases that are not of
our interest.)
Other than these restrictions, $F(x)$ can be
quite general; $F(x)$ may even be discontinuous as will be
illustrated in the next section.

Now let us consider the union $I$  of intervals, $I_n =[a_n,\,\,
b_n]$ with $0 \le a_n < b_n$, which is ordered and non
overlapping. Namely $b_n < a_{n+1}$.  If there are $l$ such
intervals, $I$ is given by $I_1 \cup I_2 \cup\cdots \cup I_l$.
Without loss of generality, one may consider $F(x)$ being
nonzero only in $x\in I$. For this setting of $F(x)$, the boundary
at $y=0$ takes in general
 the following
shapes in the $(\rho,\,\eta)$-plane. First it is clear from
(\ref{yax}) that $\eta=0$ solves  $y=0$ as $F(x)$ is chosen to be
odd. Furthermore, $\rho=0$ solves $y=0$ equation if $\eta$ does
not belong to $I$. Thus one may see that $\rho=0$-line and
$\eta=0$-line together form some part of the relevant $y=0$
boundary. The coordinate $(\rho,\, \eta)$ is defined in the region
within the first quadrant, i.e. $\rho\ge 0$  and $\eta \ge 0$.
Around each interval $I_n$, there is a larger interval $\bar{I}_n$
($I_n\subset \bar{I}_n$), in which $y=0$ boundary is given by
$\rho= g_n(\eta) \ge 0$. The intervals $\bar{I}_n$ may overlap
with each other generically. If this happens, the overlapped
intervals are merged and we may get a new set of interval
$\{\bar{I}_n\}$. This new set is constructed such that each
interval $\bar{I}_n$ is  disjoint with each other and that
$\rho=g_n(\eta)\neq 0$ within each interval. It is also possible
that the first interval may be extended to the $\rho$ axis, which
means $\rho=g_1(0)> 0$\footnote{Note that $a_1$ is zero in this
case.}. In conclusion, the $y=0$ boundary consists of $\rho$ axis
with $\rho> g_1(a_1)$, $\eta$ axis that does not belong to
$\bar{I}$, and the collection of $\rho=g_n(\eta)$ lines for $\eta
\in \bar{I}$. We illustrate the shape of boundary for $l=3$ case
in Fig.~1a by bold line.

\begin{figure}[htb]
\epsfxsize=4.4in
\vspace{.2in}
\hspace{0.7in}
\epsffile{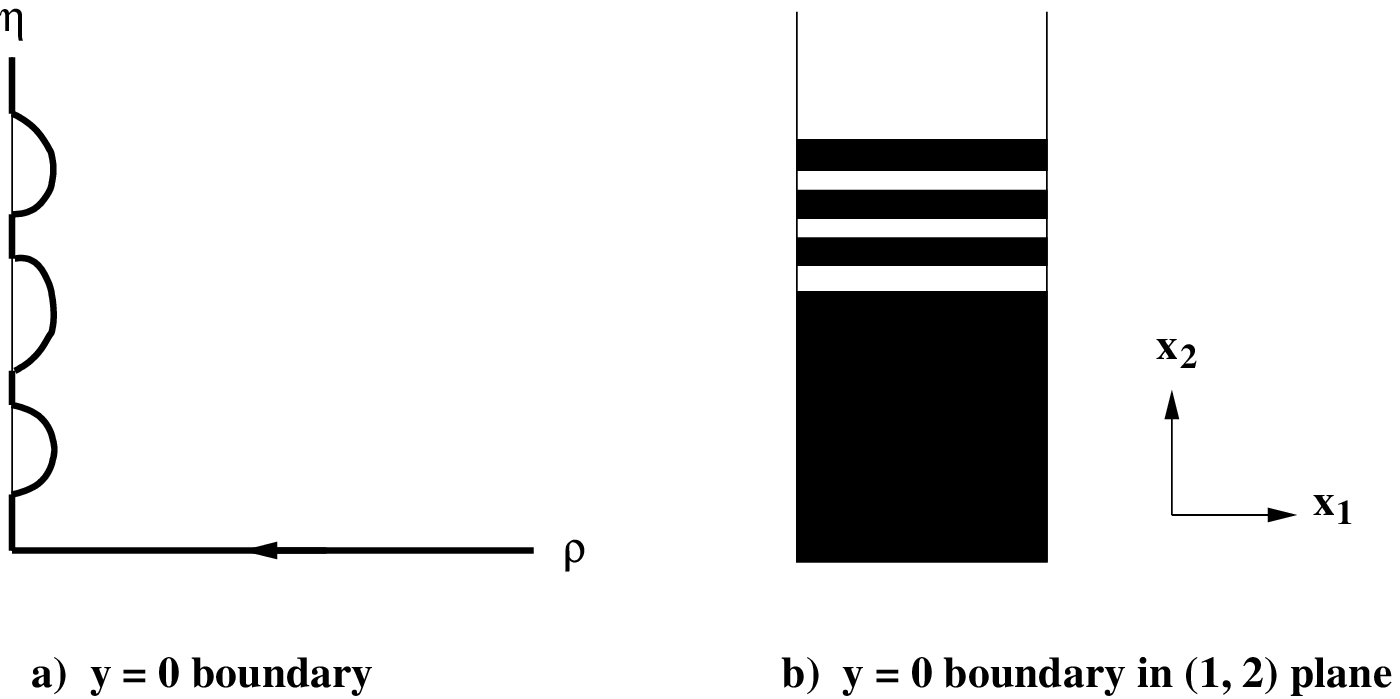}
\vspace{.2in}
\\
{\small Figure~1:~An illustration of the shape of $y=0$ boundary
in $(\rho,\,\eta)$-plane and in $(x_1,\,x_2)$ for $l=3$.}
\end{figure}

Along the boundary in $(\rho,\,\eta)$ space, that is, along the
$y=0$ level curve, $x_2$ has a monotonic behavior. For instance,
$x_2$ in Fig.~1a is monotonically increasing along the boundary in
the direction of the arrows. The proof of this follows from the
fact that the Jacobian is positive definite for $y\ge 0$, which
will be shown shortly, and that $y$ is increasing to the right of
the boundary. Then for each interval $\bar{I}_n$, one has
corresponding intervals ${\cal I}_n=[\alpha_n, \beta_n]$ in the
$x_2$ coordinate. From the monotonic behavior, the intervals
${\cal I}_n$'s are again ordered and  not overlapping with each
other. In these intervals as well as the ${\cal I}_{\rm
plane}=[-\infty, \beta_0]$ with $\beta_0=x_2 (\eta=0,
\rho=g_1(a_1))$, the finite type boundary condition is realized.
${\cal I}_{\rm plane}$ may be viewed as  M2 branes responsible for
the plane wave background.
The intervals ${\cal I}_n 
$ in the $x_2$ direction  is considered as representing   extra M2
giant gravitons. The shape of boundary in the $(x_1,x_2)$ plane is
depicted in Fig.~1b for the $l=3$ case. The number of $M2$ giant
is not given by the area of the strip but given by the area
weighted by $\rho^2$ as will be discussed in the section 6.

Let us give more details  for $l=1$ case where  $F(x)$ is non zero
only in an interval $ [a,b]$. Integrating by parts and Taylor
expanding the function $F(x)$, we can write the expression for $y$
(for small $\rho $) as: \be \label{yax1} y = -2 F(\eta) +
\frac{\rho^2}{2} \left[
\p_\eta x_2(\rho = 0)
+ 3\rho^2 g(\eta) + ...\right]
\ee
where
\[
g(\eta) = \int^{\infty}_{-\infty} dw \frac{F(\eta + w)}{|w|^5}.
\]
{}From this expression we note the following:

i. For $\rho = 0$  and $\eta \in [a,\,b]$, one can see that $ y \ne 0 $
and hence $\rho = 0$ and $\eta \in [a,\,b]$ is excluded from the boundary.

ii. The condition that $y\ge 0$ for infinitesimal $\rho$ further
requires $x'_2 \ge 0$ and one can see that the intervals
$[\eta_{min}, a]$ and $[b, \eta_{max}]$ are also excluded from the
boundary, $\eta_{min}$ and $\eta_{max}$ being the roots of
$\p_\eta x_2 = 0$ to the left of $a$ and right of $b$
respectively.

iii. The $\rho>  0$ and $\eta \notin [\eta_{min},
\eta_{max}]$ region is also not a boundary except for $\eta = 0$,
which is a trivial boundary with  boundary conditions of finite type
\[
\rho^2 = e^D \sim {\rm finite}
\]

iv. For $\rho = 0$ and $\eta \notin [\eta_{min}, \eta_{max}]$
the boundary conditions are of linear type as the function $F(\eta)$ vanishes
outside the interval $[\eta_{min}, \eta_{max}]$
\[
\rho^2 = e^D \sim y
\]

v. For $\rho>  0$ and $\eta \in [\eta_{min},
\eta_{max}]$, the boundary conditions are of finite type as $y=0$
boundary looks like a bump toward $\rho\geq 0$ region
\[
\rho^2 = e^D \sim {\rm finite}
\]
However, we always find that $\partial_y D=0$ condition is not
satisfied in this part, and the geometry has a null singularity as
will be discussed in section 5. Thus a suitable choice of the
function $F(\eta)$ (odd and positive in a finite interval) gives
the solutions which obey the almost right kind of boundary
conditions for a giant graviton except the appearance of null
singularity. In the next section we shall discuss this in an
explicit example.

By now we have not worried about smoothness of the coordinate
transformation from
 $(y, x_2)$ coordinates to $(\rho, \eta)$ coordinates.
For a well behaving solution in $(y, x_2)$ coordinates one would
require that the change of coordinates should be well-defined
throughout the region. To analyze this, consider the Jacobian of
the transformation
 $(y, x_2)\rightarrow (\rho, \eta)$
\be J = \frac{\p y}{\p \rho}\frac{\p x_2}{\p \eta} -  \frac{\p
y}{\p \eta} \frac{\p x_2}{\p \rho}. \ee Using the definitions
(\ref{defyax}) of $y$ and $x_2$ and the equation (\ref{lap}) one
finds \be
J = 
\frac{1}{\rho} \Big[ \left(\frac{\p y}{\p \rho}\right)^2 +
\left(\frac{\p y}{\p \eta}\right)^2\Big] \ee which is positive
semi-definite.

Also we have implicitly assumed that $y \ge 0$. Since $y$ is
 related to the radii of the two-sphere and five-sphere, one
 would like to show that radii of two-sphere and five-sphere
never become negative. The radii of two-sphere and five-sphere are
related to the quantity $e^{6\lambda}$ defined in the metric
(\ref{met1}). The requirement for the non-negativity of
$e^{6\lambda}$ reads from (\ref{met1}) as \be K \ge 2
\frac{y}{\rho} \frac{\p y}{\p \rho} \ee which is true if the
following inequality holds \be \label{ineq} \frac{\p y}{\p \rho}
\ge \frac{2y}{\rho}. \ee Integrating by parts the expression
(\ref{yax}) for $y$ becomes \be y  =  2 \rho^2 \eta  - \rho^2
\int^{\infty}_{-\infty} dx \frac{F(x)}{(\rho^2 +
(x-\eta)^2)^{\frac{3}{2}}}. \label{yex} \ee Taking partial
derivative with respect to $\rho$ and using the fact that $F(x)$
is an odd function, one gets \be \label{cond} \frac{\p y}{\p \rho}
=   \frac{2y}{\rho} + 3 \rho \int^{\infty}_{0} dx
\Big[\frac{\rho^2}{(\rho^2 + (x-\eta)^2)^{\frac{5}{2}}} -
\frac{\rho^2}{(\rho^2 + (x+\eta)^2)^{\frac{5}{2}}} \Big]F(x). \ee
>From this, the inequality (\ref{ineq}) follows provided $F(x) \ge
0$  for $ x \ge 0$. Hence, the non-negativity of $y$ further
requires that $F(x) \ge 0$  for $ x \ge 0$.

\section{Examples of Giant Gravitons}

To give a compelling basis to the validity of the solutions we
have described, let us analyze an explicit example more carefully.
For this purpose, we make the simplest choice of the function
$F(x)$, that is, the one with $\delta$-function distribution,
\be
F(x)= f_0 \left(\delta(x-\eta_0)-\delta(x+\eta_0)\right)\,. \ee It
is easy to calculate $y$ and $x_2$ to find \bea y&=&\rho^2\Big(
2\eta-f_0\frac{1}{[\rho^2+(\eta-\eta_0)^2]^{3\over2}}
+f_0\frac{1}{[\rho^2+(\eta+\eta_0)^2]^{3\over 2}}\Big)\quad,\nonumber\\
x_2&=&-\rho^2+2\eta^2+f_0\frac{\eta-\eta_0}{[\rho^2+(\eta-\eta_0)^2]^{3\over2}}
-f_0\frac{\eta+\eta_0}{[\rho^2+(\eta+\eta_0)^2]^{3\over 2}}\quad.
\eea
We are interested in the region of $\rho,\,\,\eta\geq 0$ in which $y\geq 0$, and
the $y=0$ boundary is of special importance as discussed before. One obvious
component of $y=0$ curve in $(\rho,\,\, \eta)$ plane is $\eta$-axis ($\rho=0$),
whereas
there may be other components by solving
\be
2\eta=f_0\Big(\frac{1}{[\rho^2+(\eta-\eta_0)^2]^{3\over 2}}-
\frac{1}{[\rho^2+(\eta+\eta_0)^2]^{3\over 2}}\Big)\quad.
\label{semicircle}
\ee
We will find in the following that there are indeed two
 additional components from this equation; one is the
$\rho\geq 0$-axis ($\eta=0$) and the other is an approximate
semi-circle (or ``bump") around
the point $(\rho,\eta)=(0,\eta_0)$ in the $\rho\geq 0$ region.
Let us call the two intersection points of this approximate
semi-circle with the $\eta$-axis, A and B,
whose positions are $(0,\eta_{\rm min})$ and $(0,\eta_{\rm max})$.
At first sight, this may seem to indicate a problem because
$y=0$-curve bifurcates at A and B into
$\rho=0$-curve and the semi-circle.
However, we will show that, for
$\eta\in [\eta_{\rm min},\eta_{\rm max}]$, we should actually
discard $\rho=0$ segment and take only semi-circle part. This
is because the semi-circle
turns out to be the correct boundary of $y\geq 0$ region,
while $\rho=0$, $\eta\in [\eta_{\rm min},\eta_{\rm max}]$
 is the boundary of $y\leq 0$ region instead.

Therefore, the whole picture of $y=0$ boundary for $y\geq 0$
region would be
\be
\{\rho\geq 0, \eta=0\}\cup\{\rho=0,
\eta\in [0,\eta_{\rm min}]\cup[\eta_{\rm max},\infty]\}
\cup\{\rho=g(\eta), \eta\in [\eta_{\rm min},\eta_{\rm max}]\}
\quad,\label{yzero}
\ee
where in the last component, $\rho=g(\eta)$ is the equation of
the (approximate) semi-circle,
whose implicit expression is given by (\ref{semicircle}).
In the previous section, we have shown that Jacobian between
$(\rho,\,\eta)$ and $(y,\, x_2)$ is positive
definite. This implies that $x_2$ is monotonic along the level
 contour of $y=0$ in $(\rho,\eta)$ plane.
Recalling the map $e^D=\rho^2$, we thus have two separate
components of $D\sim {\rm (finite)}$
boundary condition as $y\rightarrow 0$, that is, the first
and the last component in (\ref{yzero}),
while the second component in (\ref{yzero}) corresponds to
the boundary condition
$e^D\sim c \,y$ as $y\rightarrow 0$. Therefore, this solution
is describing one-band excitation
of the plane wave background, whose interpretation would be a
supergravity realization
of 1/ 2 BPS M2-brane.

We can explicitly confirm the claimed features of $y=0$
boundary of $y\geq 0$ region
in $(\rho,\eta)$ plane by
considering sufficiently large $\eta_0 \gg 1$ with $f_0\sim \CO(1)$. We may
reasonably expect these properties (e.g. the existence of
 the ``bump") not to change
for generic $\eta_0$ and $f_0$, indicating a large degeneracy
of solutions for a given
topology of one-band excitation. In solving (\ref{semicircle}),
the solution $\eta=0$
is obvious, and we look for solutions with $\eta\neq 0$. For
large $\eta_0$, the right hand side
of (\ref{semicircle}) will be small unless $\eta\sim \eta_0$,
and the left hand side tells us that $\eta$ is small, allowing
us to expand the right hand side
near $\eta=0$. This in fact gives us back
the $\eta=0$ solution. However, near $\eta\sim \eta_0$ region,
we may keep the first term
in the right hand side while the second term is neglected.
This gives us the equation
\be
\rho^2+(\eta-\eta_0)^2=\left(\frac{f_0}{2\eta}\right)^{2\over 3}\approx
\left(\frac{f_0}{2\eta_0}\right)^{2\over 3}\quad,
\ee
describing a semi-circle of radius $\left(\frac{f_0}{2\eta_0}\right)^{1\over 3}$
centered at $(0,\eta_0)$, and the constants $\eta_{\rm min/max}$ is given by
$\eta_0\pm \left(\frac{f_0}{2\eta_0}\right)^{1\over 3}$.

Around this semi-circle, we have
\be
y\approx \rho^2\left( 2\eta_0-f_0\frac{1}{[\rho^2+(\eta-\eta_0)^2]^{3\over 2}}
\right)\quad,
\ee
from which it is clear that the semi-circle bounds the $y\geq 0$
region to its right,
while $\rho=0$, $\eta\in [\eta_{\rm min},\eta_{\rm max}]$ segment
 is the boundary of a wrong region.
It is also seen that for $\eta\in [0,\eta_{\rm min}]\cup[\eta_{\rm max},\infty]$,
the $\rho=0$ line should be
taken as the boundary of $y\geq 0$ as in (\ref{yzero}).
What remains to be shown is
that $\rho\geq 0$, $\eta=0$ is the boundary of the $y\geq 0$
region to its above.
Near $\eta=0$, we have
\be
y\approx 2\rho^2\eta\left(1-\frac{3f_0\eta_0}{[\rho^2+(\eta_0)^2]^{5\over 2}}
\right)+\CO(\rho^2\eta^2)\quad.
\ee
Therefore, by taking $\eta_0$ and $f_0$ such that $(\eta_0)^4> 3f_0$, we have $y\geq 0$
above the $\eta=0$ boundary. Moreover,
the arguments in the previous section guarantee that $y\geq 0$
remains true in the whole
region bounded by (\ref{yzero}) and there is no other boundary.
We hope it is by now convincing that (\ref{yzero}) is the correct
boundary of our solution, and it represents an example of 1/2
 BPS one-band excitation of the
plane wave background.

It is straightforward to check that these solutions have the
correct boundary conditions as we approach $y\rightarrow 0$, that
is, $e^D=\rho^2$ behaves either $\sim c\,y$ or $({\rm finite})$,
except the segment $\{\rho=g(\eta), \eta\in [\eta_{\rm
min},\eta_{\rm max}]\}$ on which $\partial_y D\neq 0$. Therefore,
we have a band of conical singularity which turns out to be
null-like. In seeing this, it is useful to write the resulting
metric explicitly after we change the coordinate from
$(y,\,\,x_2)$ to $(\rho,\,\,\eta)$ as in (\ref{newmetric}). The
reason for this is because the presumably complicated function
$e^D$ of $(y,\,\,x_2)$ is simply $\rho^2$ in $(\rho,\,\,\eta)$
plane. Moreover, $(\rho,\,\,\eta)$ is preferred for another
reason; they become the usual radial coordinates of $R^6$ and
$R^3$ in the asymptotic plane-wave background, and this fact will be
useful when we calculate charges of our solutions in the later
section.


It is easy to generalize the above discussion to find explicit
multi-band solutions.
By simple superposition of one-band solutions, we have
\bea
y&=& 2\eta \rho^2
+
\rho^2\,\sum_{n=1}^{l}\Big\{-f^n_{0}\frac{1}
{[\rho^2+(\eta-\eta^n_0)^2]^{3\over2}}
+f_0^n\frac{1}{[\rho^2+(\eta+\eta^n_0)^2]^{3\over 2}}\Big\}
\quad,\nonumber\\
x_2&=&-\rho^2+2\eta^2+\sum_{n=1}^{l}\Big\{+f^n_0
\frac{\eta-\eta^n_0}{[\rho^2+(\eta-\eta^n_0)^2]^{3\over2}}
-f^n_0\frac{\eta+\eta^n_0}{[\rho^2+(\eta+\eta^n_0)^2]^{3\over 2}}
\Big\}\quad. \eea For sufficiently far separated $\eta_0^n$ and
small $f_0^n$, each $n$-th term independently generates small
semi-circle bump around $\eta\sim\eta_0^n$ in the $y=0$ boundary
of $y\geq 0$ region. The component of $\{\rho\geq 0,\,\,\eta=0\}$
and $\{\rho=0,\,\, \eta\in R^+ - \bigcup_{n=1}^{l}[\eta^n_{\rm
min},\eta^n_{\rm max}]\}$ are as before. Because the behavior near
each bump is not affected in an essential way by other bumps, the
properties of these solutions, like the nature of singularities,
should be same as in the one-band solutions, at least for generic
values of $\eta_0^n$ and $f_0^n$. As it will become clear in the later
section, the charges for multi-band excitations turn out to be
additive, conforming to the naive expectation.

We would like to mention that the above solutions obtained
from $\delta$ function distribution
of $F(x)$ 
are mere examples of vast possible choices of $F(x)$. For
instance, we may try a step function: $F(x)=f_0$ for $x\in
[a,\,\,b]$, $-f_0$ for $x\in [-b,\,\,-a]$, and otherwise zero. We
have analyzed this case to find one-band excitations for
reasonable values of $f_0,a$ and $b$, and also expect multi-band
solutions by superposition of them. We do not believe that
different choices of $F(x)$ are related to each other by
coordinate reparametrization, and this suggests a huge degeneracy
of solutions parameterized by a single function $F(x)$. The
physical interpretation of this function will become clear in
section 6 when we discuss charges of the solutions.

\section{Null Singularity}

Our solutions presented in (\ref{master}) are singular at
$y=0$ if $\eta\  \in\ \bar{I}$. In this region of $y=0$ with
$\eta\  \in\ I$, the finiteness condition of $D$ is
satisfied but $\partial_y D=0$ is not satisfied.  This can be
seen as follows.  Note that
\be
e^{6\lambda}= {y(\rho^2 - y \partial_y \rho^2)\over \partial_y \rho^2 }
= {y \rho K \over 2 \partial_\rho y} -y^2
\ee
where the first equality is from (\ref{met}).
{}From the second equality, one may see that
$\partial_y D = 2\partial_y \ln \rho=0$ condition is
equivalent to $\partial_\rho y=0$. 
However one may see
from (\ref{cond}) that $\partial_\rho y$ is  finite (and positive)
 for a finite $\rho >0$ and for any nontrivial $F$ ($F(x) \ge 0$ when
$x\ge 0$). Thus the boundary condition is not satisfied for $\eta
\in \bar{I}$ leading to the above mentioned singularities.
At the singularities, one finds that   $e^{2\lambda}$ as well as
the radii of both the  two and five spheres are vanishing. These
are null-like conical singularities and not much different from
those of extremal  branes, whose  horizon is singular in general.

Before we discuss the nature of the singularities further, we like
to argue that these singularities are in fact unavoidable in our
framework for the geometries of 1/2 BPS giant gravitons asymptotic to
the plane wave geometry.
 The most general solution of the Laplace
equation that is asymptotic to the plane wave geometry is given by
\be V= \rho^2 \eta -{2\over 3}\eta^3 + \int^\infty_{-\infty}dx
\int_0^\infty ds s \int^{2\pi}_0 d\theta {F(s, x)\over \sqrt{
\rho^2 +s^2 -2s\rho \cos\theta +(\eta-x)^2} }
\quad,\label{general}\ee where the source $F(\rho, \eta)$ should
be distributed outside of $y\geq 0$ region in
($\rho$,$\eta$)-plane. In writing the last part $V_1$ in
(\ref{general}), we used the fact that $V_1$ satisfies the three
dimensional Laplace equation with the rotational symmetry around
$\eta$ axis, and also the boundary conditions at infinity for the
asymptotically plane wave geometries.  Let $\rho_s$ be the maximum
radius of the
 nonvanishing  source distribution occurring at
$\eta=\eta_s$. After a short calculation, we find
\be
\frac{\partial y}{\partial\rho}=\frac{2y}{\rho}+
\int^\infty_{-\infty}dx
\int_0^\infty ds s \int^{2\pi}_0 d\theta \left[\frac{-2s\cos\theta F(s,x)}{\rho Z^{3\over 2}}
+\frac{3(\rho^2+s^2-2s\rho\cos\theta) F(s,x)}{Z^{5\over 2}}\right]\quad,
\ee
where $Z=\rho^2+s^2-2s\rho\cos\theta+(\eta-x)^2$. As before, requiring
$F(s,x)$ be odd in $x$ and positive definite for $x\geq 0$ for the right shape
of the $y=0$ boundary as in Fig.1a, it can be easily proved that the second term
in the above is always positive. This implies that $y$ remains positive in the bulk, and
we have conical singularities on the ``bump" of $y=0$-boundary,
exactly same as in the line charge case.

Indeed, the potential (\ref{general}) is equivalent to the one from a line charge density
for outside observers with $\rho >\rho_s$.
Using the definition of $K_0$, the $V_1$ part may be rewritten
as \be V_1={1\over 2\pi} \int^\infty_{-\infty}dk
 \int^\infty_{-\infty}dx
\int_0^\infty ds\, s \,\,
 e^{ik(\eta-x)}  {F(s, x)} \int^{2\pi}_0 d\theta   K_0 (kZ(\theta))
\label{v1} \ee where $Z^2(\theta)= s^2 +\rho^2 -2s\rho\cos\theta$.
We then perform the $\theta$ integration. Using  the formula \be
{1\over 2\pi} \int^{2\pi}_0 d\theta   K_0 (kZ(\theta)) = I_0 (ks)
K_0 (k\rho)\, \ \ \ \ \ \  \ \  {\rm for}\  s \le \rho\,, \ee one
may rewrite (\ref{v1}) as \be V_1= \int^\infty_{-\infty} dx
{\bar{F}(x)\over \sqrt{ \rho^2+(\eta-x)^2} } \, \ \ \ \ \ \  \ \
{\rm for}\  \rho_s \le \rho\,, \ee where we define \be \bar{F}(x)
=2\pi\int_{-\infty}^{\infty} dk \int_{-\infty}^{\infty} dy \int_0^\infty ds\, s  e^{ik(x-y)} I_0(k s) F(s,y)
\,. \ee Therefore, for the charge distribution localized around the $\eta$
axis with the cylindrical symmetry, the system may be equivalently
described as a line charge density at $\eta$ axis for the observer
outside the maximum of  $\rho$ ($\rho_s$) of the charge
distribution. If $F(s,x)$ is odd under $x$, $\bar F(x)$ is also odd.
The nice properties derived in the previous paragraph can be
seen more easily in this context because
we showed in section 2 that they hold for line charge distributions.

We think there is in fact a stronger argument for the singularity on
the ``bump"
of $y=0$ boundary. Quite generally, suppose we have a solution with the necessary
boundary condition on the $y=0$ curve satisfied, and imagine focusing on the
edge point where the boundary condition changes from the linear type to the finite type.
Note that for this purpose, it will be sufficient to assume that we can
go to the $(\rho,\eta)$ description only locally around this point, and the conclusion
we draw from this is universal in this sense.
In $(\rho,\eta)$-plane, the $y=0$ curve leaves the $\eta$-axis at this point, and let us consider
the expansion of the potential around it.
(This kind of point should exist for the giant gravitons unless
$y=0$ line stops at the $\eta$ axis.) The general argument shows that
this kind of point should occur away from the charge distribution.

Taylor expansion around the point $(0,\eta_0)$
should give us polynomial type of
solutions of the Laplace equation, which can be constructed order by order
by the formula
\be
V=\sum_{n\geq 0} F_n(\rho^2)\tilde\eta^n\quad,
\ee
with the recursion relation
\be
F_{n+2}(x)=-\frac{4}{(n+2)(n+1)}\left(x F''_n(x)+F'_n(x)\right)\quad,
\ee
where $\tilde{\eta}=\eta-\eta_0$.
The sum 
is actually finite for each independent solution.
To describe the $y=0$ curve leaving $\eta$-axis, the lowest
order expansion of the potential $V$ looks like
\be
V= c_0 + c_1 \tilde\eta + c_2 \Big(\rho^2\tilde\eta -{2\over 3}
{\tilde\eta}^3\Big)
+ c_3 \Big(\rho^4 - 
8 \rho^2 {\tilde\eta}^2+ {8\over 3}{\tilde\eta}^4 \Big) + \cdots
\ee
Then
$y=\rho\partial_\rho V$ becomes
\be
y= 2\rho^2 (c_2 \tilde{\eta} + 2 c_3 \rho^2 - 8c_3 {\tilde\eta}^2 + \cdots)
\ee
where $c_2$ should be nonzero for
the boundary condition of linear type.
One may see that $y=0$ curve is described by $\rho=0$ and the curve of
\be
{c_2}\tilde\eta= - 2 c_3 \rho^2 + \cdots
\ee
The evaluation of $\partial_\rho y$ leads to
\be
\partial_\rho y =4\rho (c_2\tilde\eta + 4 c_3 \rho^2 -
8c_3 {\tilde\eta}^2
\cdots)
\ee
Thus along $y=0$ with $\rho\neq 0$, we see that $\partial_\rho y \neq 0$
in general. In case  $c_3=0$, one has to consider the higher order
terms
but the conclusion is not changing, unless $V(\rho,\eta)$ exactly factorizes
into $F(\rho^2)G(\eta)$ where we find no relevant solution.

As mentioned earlier, the only remaining possibility is that
$y=0$ line stops at somewhere in the $\eta$ axis. This is
possible as $y$ works as a coordinate. A possible example arises
for the case of $F(x) \le 0$ for $x \ge 0$. In this case, at the boundary
of $\eta$ axis excluded by the $y=0$ line, the radii of two and five
spheres become finite while $\rho$ goes to zero. To make a smooth
geometry, one has to utilize $x_1$ circle but there is no way
to match the period of the $x_1$ circle except for some special
cases that are not of
our interest.

This completes our argument for the non-existence of {\it regular} 1/2 BPS
giant solutions
that are asymptotic to the plane wave geometries.
Therefore in the supergravity description of the ansatz we are considering,
the occurrence of singularities seems
inevitable. There are many examples of singular solutions in
supergravity theories. As mentioned earlier, the singularities of our
solutions are null-like and rather similar to the null
singularities occurring
in the supergravity description of extremal D-branes.
In case of extremal D-branes, we view the D-branes as  extra
fundamental degrees of freedom corresponding to sources of supergravity fields.
Without the source terms describing the
D-brane dynamics, the supergravity description would miss some
part of the degrees of freedom leading to singularities. Indeed the structure of singularity matches precisely with the Born Infeld type source term
for D-branes.
Whether M2-brane description \cite{Bak1} of the giant graviton
dynamics
matches with the singularity structure of the above
solution is of particular interest. Further studies are required in
this regard.

\section{Charges}

In this section, we like to identify the charge content of the
1/2 BPS system. This will be essential for the later comparison
with the 1/2 BPS states of the matrix model in the later section. First consider the lightcone momentum, $p^+$ the charge associated with the translation
along $x_1$ direction, which counts the number of D0 branes in the
IIA string theory.

Since the
solutions are asymptotic to the plane wave, the usual ADM definition
of energy and momentum will not work for our case.  In Ref.~\cite{Har}, a method of finding
the conserved quantities for the background with translational isometry was discussed.
Taking $x_1$ as a relevant isometry
direction, one may get the expression for the charge
straightforwardly. Here we would like to present an alternative
method for computing the charge, which gives precisely the same results.
We first go to the IIA string description  compactifying the eleven dimensional theory
on the circle along $x_1$ direction. The ten dimensional metric in the string frame
is related to the eleven dimensional metric by
\be
ds^2_{11}= e^{-{2\phi\over 3}} ds^2_{\rm 10} +  e^{{4\phi\over 3}} (dx_1^2 + C_\mu dy^\mu)^2
\ee
where $\phi$ is the dilaton field, $y^\mu$ is the coordinate for the ten dimensions
and $C_\mu$ is identified with the R-R one form potential. 
The integral of
$*dC$ over the eight sphere at infinity corresponds to the charge.
Working with the following general form of the metric,
\be
ds^2= - h_t dt^2   + 2w \, dt dx_1+ h_x dx_1^2 + h_{ij} dy^i dy^j\,,
\ee
the IIA metric becomes
\be
ds_{10}^2=  \sqrt{h_x}\left( - {L\phantom{i}\over h_x}\, dt^2  + h_{ij} dy^i dy^j\right)\,,
\ee
where we introduce $L= w^2 + h_t h_x$, and
$e^{{4\phi\over 3}}= h_x$. The only non-vanishing component of $C_\mu$ is
$C_t= w/h_x$. The expression for the charge  reads
\be
Q={1\over 2k^2_{11}}
\int * dC = {1\over 2k^2_{11}} \int_{y=\infty}
\sqrt{{\rm det}h }
\left[-\partial_i h_x + h_x {\partial_i w/
w}\right]
(1+ h_t h_x/w^2)^{-1/2} h^{ij}n_j r^8 d\Omega
\ee
where $r^2= y^i y^i$ ($\sim \rho^2 + \eta^2$ for large $r$),
the integration is over unit eight sphere 
and we restore 
$2k^2_{11}$ dependence. 
The vector
 $n_i$ that is normal
to the eight sphere
 is defined as  $n_i n_i=1$.

Let us consider  the  solutions with general $F(x)$, for which
the metric
components at large $r$ behave as,
\bea
h_t = O(r^2)\,, \ \ \  w= - 2 +O(1/r^5)\,, \ \ \  h_{ij}= \delta_{ij}
+ O(1/r^5)\,,
\eea
and
\be
h_x = {2y \rho\over e^{4\lambda} \partial_\rho y}
\left[{\rho\partial y\over y \partial \rho}-1\right]=
{3p\over \eta r^5} + {15 q\over  r^7} + \cdots 
\label{asym}
\ee
where
 $p$ and $q$ are defined by
\bea
p ={1\over 2} \int^{\infty}_{-\infty}
dx F(x)\,, \ \ \   
q ={1\over 2} \int^{\infty}_{-\infty} dx x F(x)\,.
\eea
The first relation in (\ref{asym}) follows directly  from the metric (\ref{met1}).  The
asymptotic  expansion of $h_x$ may be obtained
from the
 expansion of $y$ in (\ref{yex}),
\be
y  =  2 \rho^2 \eta \left(1- {2p\over \eta r^3} - {6q\over  r^5}
 +\cdots\right)\,.
\ee

{}From the asymptotic behavior of the metric components,
it is clear that $Q$ diverges
once $p$ is non-vanishing. This is one reason why we choose the function $F(x)$
to be  odd.   We do not claim that the
solution is unphysical  but only that the
charge $p^+$ is not well-defined for nonzero $p$.
For our choice of  odd function $F(x)$, the charge $Q$
becomes
\be
Q= {105 \,\omega_8\over 2k^2_{11}}\,\,  q\,.
\ee
where $\omega_8$ denotes the volume of the eight sphere
of unit size.

Now let us discuss the number of the M2 giants. As discussed in
Ref.~\cite{lin}, this  may be obtained by evaluating
the flux $* G_{(4)}$ through seven cycle ($\Sigma_7$) of
the five sphere fibered over two-surface which ends on the $y=0$
region where  the five sphere shrinks. This leads to
\be
N_2 \sim {1\over \omega_5} \int_{\Sigma_7} * G_{(4)}=\int_{{\cal D}} dx_1
dx_2 2 e^D|_{y=0}\,,
\ee
where ${\cal D}$ is the region in the 12 plane where $S^2$
shrinks.
The evaluation of this number for general $F(x)$ appears
to be complicated. But when $F(x)$ is small enough and non-vanishing only for
large $x$, $N_2$ is given simply by
\be
\Delta N_2 \sim \int_0^\infty dx F(x)\,,
\label{number}
\ee
where $\Delta N_2 $ counts only the extra M2 branes. When $F$ is small enough,
the main contributions
comes from the intervals $I$, for which $2 e^D \sim F(\eta)/\eta$ and
$dx_2\sim 4\eta d\eta$.
Note that the argument $x$ in $F(x)$ is related to $\eta$ as one
may see for instance in the expression  (\ref{yex}). Since $\eta$ is
closely related to the radius of the two-sphere,
the natural interpretation of the variable $x$ would be the size
of  giant M2-brane.  In comparison with the BMN matrix model in the next section,
this interpretation  will be clearer.  Since the integral
$\int_0^\infty F(x)\, dx$ counts the number of M2-giants, $F(x)$ may naturally be
interpreted as the number density of the giant
M2 with respect to its radius $x$. Since $p^+$ is proportional to
$\int_0^\infty F(x) x \,dx$, one can see that each M2 should carry
the $p^+$ charge proportional to the size $x$. Later in the BMN matrix
model, one may indeed confirm this prediction.

\section{Correspondence to the BMN Matrix Model}

The dynamics of the plane wave geometry has a matrix model
description. The matrix model may be obtained by the standard
method of regularizing the supermembrane action in the
maximally supersymmetric
plane
wave background \cite{bmn,Das}.  The $x^-$ ($x_1$ in the previous section) is the
isometry direction and one is compactifying this direction on
a circle of radius $R$ and $x^+$ ($t$ in the previous section)
serves as the lightcone
time direction. The conserved momentum of the supermembrane
along the $x^-$ direction is quantized
$p^+=p_-=N/R$.
In the IIA description, the integer $N$ counts
the number of D0 branes. The matrix variables are given by
$N\times N$ Hermitian matrices as they represent
the strings connecting $N$ D0 branes.

The BPS states  of this BMN matrix model have been
studied in detail\cite{bmn,Das,Bak1}.
The 1/2 BPS states corresponding to the giant
gravitons are particularly simple.  The bosonic part has  nine spatial
directions described by $N\times N$ hermitian matrices and
has $SO(3)\times SO(6)$ R-symmetry.

The 1/2 BPS states are governed by
\be
[X_a, X_b]= i\epsilon_{abc}\, X_c
\ee
where the indices a, b, c  run over 1, 2, 3 and the other
directions remain unexcited.

The construction of the general 1/2 BPS states is simple. Let
us first take an
$n$ dimensional irreducible
representation of the $SU(2)$ algebra where $n$ is
a positive integer.  The Casimir $L^2= X^2+ Y^2+Z^2$ is given by
$j(j+1)=(n^2-1)/4$ where $j=(n-1)/2$.  Thus the radius of the fuzzy sphere
corresponding to the giant graviton is given by
\be
R_g= {\sqrt{n^2-1}\over 2}\,.
\ee
Then  generic  1/2 BPS states are given by
any combination  of the irreducible representations,
\be
T=[n_1]\oplus [n_2] \oplus [n_3]+ \cdots \oplus [n_k]
\label{young}
\ee
where $\sum^k_{i=1} n_i=N$
and we take $n_1 \le n_2 \le \cdots \le n_k$.
This may be thought of as representations of two dimensional Young diagrams with
total $N$ boxes.
Since the gauge equivalent representations are not
distinguished, we only care about the
combinations and the ordering of the dimensions
does not matter.
The 1/2 BPS state $T$ may be presented alternatively
by
\be
T= \sum^\infty_{n=1}\oplus f_n [n]
\ee
where $f_n$ is the multiplicity of the $n$ dimensional
irreducible representations. The sum  $\sum^\infty_{n=1}f_n$
counts the number
of total giant gravitons.

The dimension $n$ of the irreducible
 representation is related to
the radius of the fuzzy sphere and it is natural to relate it
to the argument $x$ of $F(x)$, the function characterizing our solutions. One may then see the correspondence
\be
\int_0^\infty dx x F(x) \ \ \leftrightarrow \ \  \sum^\infty_{n=1} n f_n\,,
\ee
which counts the number $N$ of the D0 branes. In addition,  one has
\be
\int_0^\infty dx  F(x) \ \ \leftrightarrow \ \
\sum^\infty_{n=1}  f_n\,,
\ee
which counts the number of the giant gravitons. Therefore, the physical
meaning of $F(x)$ is the density of SU(2)-representations of dimension $x$
in large $x$ continuum limit.

This completes the description of the correspondence. One can see that the
data for the giant gravitons in the supergravity description is encoded
in the arbitrary function $F(x)$ for the interval $x\ge 0$. The
arbitrariness of $F(x)$ corresponds to the arbitrariness of the number of
giant graviton for a given dimension $n$. Hence one may say that the
 correspondence works for  any possible states. Namely any
1/2 BPS states of the matrix theory specified by
$\{f_1,f_2, f_3 \cdots\}$ has a corresponding supergravity
data specified by the function  $F(x)$ for $x\ge 0$:
\be
\{f_1,f_2, f_3 \cdots\}
\ \ \leftrightarrow \ \
F(x)\ {\rm for}\  x\ge 0\,.
\ee
Thus the counting of states for a fixed charge $p^+$ is well
defined in the matrix model description. For the case of the
supergravity, the function $F(x)$ is continuous and the volume of the
fluctuation of $F(x)$ for the fixed $p^+ =\int_0^\infty dx x F(x)$
will become zero because we are dealing with infinite dimensional
space.
The situation here is rather similar to the counting
of the fluctuation of the supertube moduli space\cite{Bak3}. The trouble follows
from the continuum nature of the description;
What we need
is a quantization or regularization.

Such a regularization is already achieved in the matrix model
description. The state are even protected by the quantum corrections
due to the enough number of supersymmetries.
The counting of states for a fixed $p^+= N$ is simple.
It is the counting of all possible combination of the
states $\{f_1, f_2,f_3,\cdots\}$ while fixing the sum
$\sum^\infty_{n=1}n f_n$. This problem may be mapped to the problem of
the bosonic oscillators with Hamiltonian,
\be
H= \sum^\infty _{n=1} {n a_n^\dagger a_n}
\ee
where $a_n$ satisfies
 the commutation relation $[a_n, a^\dagger_m]=\delta_{nm}$
while all the remaining commutators vanish.
Then  $f_n$ corresponds to the occupation number of the n-th oscillator
$a_n$.
The partition
function is given by
\be
Z(w)= {\rm tr} e^{-\beta H}=\prod^\infty _{n=1} (1-w^n)^{-1}
\ee
where $w=e^{-\beta}$.
It  is related to the Dedekind eta function
\be
\eta (\tau) = e^{i\pi\tau /12} \prod^\infty_{n=1} (1-e^{2\pi in\tau})
\ee
The degeneracy of the state for the fixed $N$ is obtained by
\be
d_N= \oint {Z(w)\over w^{N+1}} {dw\over 2\pi i}\,.
\ee
Evaluating this  using the  saddle point method,
one obtains
\be
d_N \sim  {1\over 4\sqrt{3}\,N} e^{\pi\,\sqrt{{2\over 3}N} }\,,
\ee
for the large $N$.
 The entropy is then given by
\be
 S_N=\ln d_N \sim \pi \sqrt{{2\over 3}N}\,.
\ee
The counting problem here basically corresponds to finding
the possible ways  of partitioning
$N$ into positive  integers.

Since our starting representation in (\ref{young}) is
the representation of the 2d Young diagram, the above chiral boson
problem is equivalent  to the problem of the 2d Young diagram.
Or the above (chiral) boson may be mapped into the NS sector of
a complex fermion through the well known bosonization
technique.
Then the states of NS sector of the fermion is
in one to one correspondence with the 2d Young diagram \cite{Okoun}.
This 2d Young diagram may be mapped into the c=1 matrix model
with a harmonic oscillator potential, which is related to
the $N$ noninteracting  fermions under a harmonic oscillator
potential \cite{ber}. But from the data of the supergravity, this relation
to the fermion seems unmotivated. Namely, we do not find any
shape of Fermi liquid in ($x_1$, $x_2$) plane of the supergravity
description unlike case of the 1/2 BPS geometry of  the
type IIB string theory \cite{lin}.

\section{Conclusion}

The giant gravitons
in various $AdS_n\times S^m$ spaces as well as in the plane wave
limit have been studied  in the context of AdS/CFT correspondence.
Recent progress on constructing full back-reacted supergravity geometries
of $1/2$ BPS giant gravitons, initiated in Ref.\cite{lin},
gave us much insight
on the correspondence including
the emergent fermionic system in the type IIB string theory
in $AdS_5\times S^5$.
However, given the situation where smooth solutions
in the asymptotic plane-wave background in 11D supergravity have not
been found,
it is an important problem to find solutions in the recent
framework of LLM in Ref.\cite{lin}. In this work,
we performed a careful analysis for this problem, starting from
the linear Laplace equation that Ref.~\cite{lin} proposed.
We found a class of solutions that involve an arbitrary
function $F(x)$ whose interpretation has a nice correspondence to the
BMN matrix model,
that is, it is the number density of spherical M2-branes in the
transverse $R^3$-space with
respect to its radius $x$. However, we found that null-like conical
singularities
are present along the M2-branes, and tried to argue that this is a
necessary consequence
of the given ansatz.

There is one possible scenario to understand the nature of troubles
that we encounter in this paper. Let us first note that
analogous problems arise in the study of charged rotating
BPS black holes
in four dimensional N=2 supergravity theories. All the known solutions
have naked singularities. The dual descriptions by strings or D
branes are well understood and counting degeneracy is
straightforward\cite{Dab1}. Thus the corresponding rotating
supergravity black hole solutions should exist.  In Ref.~\cite{Gaunt1},
it is shown that  nonsingular BPS solutions may be constructed by
the addition of massive Kaluza Klein fields in such a way that the
solutions decompactify near the core to five dimensional black hole
solutions with regular horizon.
Only  asymptotically,
they approach  four dimensional geometries
(times circle).
If one tries to eliminate the massive Kaluza Klein modes by smearing
over the circle, the geometries become the known four dimensional
solutions that are
 naked singular.
If this  phenomenon happens in our case too, it implies that
the assumption of the isometry along $x_1$ circle is too
strict. Namely, dropping the ansatz of the isometry and
solving the full three dimensional
 Toda equation, there may be  a possibility of
resolving the singularities.  In  this 
case, our solutions
clearly
 have their physical meaning and  validity  once away
from the singularities; Only the core
singularities  need to be remedied  by uplifting to  higher dimensions.
Of course  our problem  is then
naturally extended  to the problem of finding 1/2  BPS
giant gravitons
asymptotic to $AdS_4\times S^7$ or
 $AdS_7\times S^4$ geometries.
Further studies are required in these directions.


\vskip 0.5cm

\noindent{\large\bf Acknowledgment} \vskip 0.5cm
 { We would like
to thank Kimyeong Lee and Soo-Jong Rey for helpful
discussions. H.-U.Y. thanks
Danilo Diaz and Oleg Lunin for discussions on singularity during
2005 ICTP spring school at Trieste, and also Jin-Ho Cho, Seok Kim
and Joris Raeymaekers for discussions.
D.B. also likes to thank the warm
hospitality of KIAS where part of this work was done.
D.B. is supported in part
by KOSEF ABRL R14-2003-012-01002-0 and KOSEF R01-2003-000-10319-0.
H.-U.Y. is partly supported by grant No.R01-2003-000-10391-0 from
the Basic Research Program of the Korea Science \& Engineering
Foundation. The work of S.S. is supported by KRF grant No.
R14-2003-012-01001-0.}



\begin{thebibliography}{99}

\bibitem{lin} H.~Lin, O.~Lunin and J.~Maldacena,
 ``Bubbling AdS space and 1/2 BPS geometries,''
 JHEP {\bf 0410} (2004) 025
 [arXiv:hep-th/0409174].

\bibitem{Lunin:2001jy}
  O.~Lunin and S.~D.~Mathur,
  ``AdS/CFT duality and the black hole information paradox,''
  Nucl.\ Phys.\ B {\bf 623}, 342 (2002)
  [arXiv:hep-th/0109154].\\
  O.~Lunin, J.~Maldacena and L.~Maoz,
  ``Gravity solutions for the D1-D5 system with angular momentum,''
  [arXiv:hep-th/0212210].





\bibitem{Martelli:2004xq}
  D.~Martelli and J.~F.~Morales,
  ``Bubbling AdS(3),''
  JHEP {\bf 0502}, 048 (2005)
  [arXiv:hep-th/0412136].

\bibitem{Liu:2004hy}
  J.~T.~Liu and D.~Vaman,
  ``Bubbling 1/2 BPS solutions of minimal six-dimensional supergravity,''
  [arXiv:hep-th/0412242].\\
  J.~T.~Liu, D.~Vaman and W.~Y.~Wen,
  ``Bubbling 1/4 BPS solutions in type IIB and supergravity reductions on S**n
  x S**n,''
  [arXiv:hep-th/0412043].


\bibitem{Bena:2004qv}
  I.~Bena and D.~J.~Smith,
  ``Towards the solution to the giant graviton puzzle,''
  Phys.\ Rev.\ D {\bf 71}, 025005 (2005)
  [arXiv:hep-th/0401173].\\
  A.~Buchel,
  ``Coarse-graining 1/2 BPS geometries of type IIB supergravity,''
  [arXiv:hep-th/0409271].\\
  N.~V.~Suryanarayana,
  ``Half-BPS giants, free fermions and microstates of superstars,''
  [arXiv:hep-th/0411145].\\
  M.~M.~Caldarelli, D.~Klemm and P.~J.~Silva,
  ``Chronology protection in anti-de Sitter,''
  [arXiv:hep-th/0411203].\\
  Z.~W.~Chong, H.~Lu and C.~N.~Pope,
  ``BPS geometries and AdS bubbles,''
  [arXiv:hep-th/0412221].\\
  M.~M.~Sheikh-Jabbari and M.~Torabian,
  ``Classification of all 1/2 BPS solutions of the tiny graviton matrix
  theory,''
  [arXiv:hep-th/0501001].\\
  P.~Horava and P.~G.~Shepard,
  ``Topology changing transitions in bubbling geometries,''
  JHEP {\bf 0502}, 063 (2005)
  [arXiv:hep-th/0502127].\\
  Y.~Takayama and K.~Yoshida,
  ``Bubbling 1/2 BPS geometries and Penrose limits,''
  [arXiv:hep-th/0503057].

\bibitem{malda}
  J.~M.~Maldacena,
  ``The large N limit of superconformal field theories and supergravity,''
  Adv.\ Theor.\ Math.\ Phys.\  {\bf 2}, 231 (1998)
  [Int.\ J.\ Theor.\ Phys.\  {\bf 38}, 1113 (1999)]
  [arXiv:hep-th/9711200].\\
E.~Witten,
 ``Anti-de Sitter space and holography,''
 Adv.\ Theor.\ Math.\ Phys.\  {\bf 2} (1998) 253
 [arXiv:hep-th/9802150].\\
S.~S.~Gubser, I.~R.~Klebanov and A.~M.~Polyakov,
``Gauge theory correlators from non-critical string theory,''
Phys.\ Lett.\ B {\bf 428} (1998) 105 [arXiv:hep-th/9802109].

\bibitem{bmn} D.~Berenstein, J.~M.~Maldacena and H.~Nastase,
``Strings in flat space and pp waves from N = 4 super Yang
Mills,'' JHEP {\bf 0204} (2002) 013 [arXiv:hep-th/0202021].

\bibitem{mst} J.~McGreevy, L.~Susskind and N.~Toumbas,
``Invasion of the giant gravitons from anti-de Sitter space,''
JHEP {\bf 0006} (2000) 008 [arXiv:hep-th/0003075].





\bibitem{ber} D.~Berenstein,
 ``A toy model for the AdS/CFT correspondence,''
JHEP {\bf 0407} (2004) 018
[arXiv:hep-th/0403110].\\
S.~Corley, A.~Jevicki and S.~Ramgoolam,
``Exact correlators of giant gravitons from dual N = 4 SYM
theory,'' Adv.\ Theor.\ Math.\ Phys.\  {\bf 5} (2002) 809
[arXiv:hep-th/0111222].

\bibitem{Caldarelli:2004ig}
  M.~M.~Caldarelli and P.~J.~Silva,
  ``Giant gravitons in AdS/CFT. I: Matrix model and back reaction,''
  JHEP {\bf 0408}, 029 (2004)
  [arXiv:hep-th/0406096].



\bibitem{Lee:2004kv}
  H.~K.~Lee, T.~McLoughlin and X.~k.~Wu,
  ``Gauge / gravity duality for interactions of spherical membranes in
  11-dimensional pp-wave,''
  [arXiv:hep-th/0409264].

\bibitem{Mandal:2005wv}
  G.~Mandal,
  ``Fermions from half-BPS supergravity,''
  [arXiv:hep-th/0502104].

\bibitem{Ward}
  R.~S.~Ward,
  ``Einstein-Weyl Spaces And SU(Infinity) Toda Fields,''
  Class.\ Quant.\ Grav.\  {\bf 7} (1990) L95.



\bibitem{Blau}
 M.~Blau, J.~Figueroa-O'Farrill, C.~Hull and G.~Papadopoulos,
 ``Penrose limits and maximal supersymmetry,''
 Class.\ Quant.\ Grav.\  {\bf 19} (2002) L87
 [arXiv:hep-th/0201081].

\bibitem{Gauntlett:2002fz}
  J.~P.~Gauntlett and S.~Pakis,
  ``The geometry of D = 11 Killing spinors,''
  JHEP {\bf 0304}, 039 (2003)
  [arXiv:hep-th/0212008].\\
  J.~P.~Gauntlett, J.~B.~Gutowski and S.~Pakis,
  ``The geometry of D = 11 null Killing spinors,''
  JHEP {\bf 0312}, 049 (2003)
  [arXiv:hep-th/0311112].\\
J.~P.~Gauntlett, D.~Martelli, S.~Pakis and D.~Waldram,
``G-structures and wrapped NS5-branes,'' Commun.\ Math.\ Phys.\
{\bf 247}, 421 (2004) [arXiv:hep-th/0205050].


\bibitem{Har}
  T.~Harmark and N.~A.~Obers,
  ``General definition of gravitational tension,''
  JHEP {\bf 0405}, 043 (2004)
  [arXiv:hep-th/0403103].

\bibitem{Das}
  K.~Dasgupta, M.~M.~Sheikh-Jabbari and M.~Van Raamsdonk,
  ``Matrix perturbation theory for M-theory on a PP-wave,''
  JHEP {\bf 0205}, 056 (2002)
  [arXiv:hep-th/0205185].


\bibitem{Bak1}
  D.~Bak,
  ``Supersymmetric branes in PP wave background,''
  Phys.\ Rev.\ D {\bf 67}, 045017 (2003)
  [arXiv:hep-th/0204033].\\
  D.~Bak, S.~Kim and K.~Lee,
  ``All higher genus BPS membranes in the plane wave background,''
  [arXiv:hep-th/0501202].

\bibitem{Bak3}
  D.~Bak, Y.~Hyakutake, S.~Kim and N.~Ohta,
  ``A geometric look on the microstates of supertubes,''
  [arXiv:hep-th/0407253].\\
  D.~Bak, Y.~Hyakutake and N.~Ohta,
  ``Phase moduli space of supertubes,''
  Nucl.\ Phys.\ B {\bf 696}, 251 (2004)
  [arXiv:hep-th/0404104].

\bibitem{Okoun}
  A.~Okounkov, N.~Reshetikhin and C.~Vafa,
  ``Quantum Calabi-Yau and classical crystals,''
  [arXiv:hep-th/0309208].



\bibitem{Dab1}
  A.~Dabholkar, J.~P.~Gauntlett, J.~A.~Harvey and D.~Waldram,
  ``Strings as Solitons \& Black Holes as Strings,''
  Nucl.\ Phys.\ B {\bf 474}, 85 (1996)
  [arXiv:hep-th/9511053].






\bibitem{Gaunt1}
  J.~P.~Gauntlett and T.~Tada,
  ``Entropy of four-dimensional rotating BPS dyons,''
  Phys.\ Rev.\ D {\bf 55}, 1707 (1997)
  [arXiv:hep-th/9610046].




\end{thebibliography}
\end{document}